\documentclass[conference,a4paper]{IEEEtran}
\usepackage{}

\usepackage{amsfonts}
\usepackage{amssymb}
\usepackage{mathrsfs}
\usepackage{amsmath}
\usepackage{cite}
\usepackage{mathrsfs}
\usepackage[ruled,vlined]{algorithm2e}
\usepackage{pgf}
\usepackage{tikz}
\usetikzlibrary{arrows,automata}
\usepackage[latin1]{inputenc}
\usepackage{verbatim}
\makeatletter

\newcommand{\Rmnum}[1]{\expandafter\@slowromancap\romannumeral #1@}
\makeatother

\newtheorem{thm}{Theorem}
\newtheorem{lemma}[thm]{Lemma}
\newtheorem{eg}{Example}

\newtheorem{defn}{Definition}

\newtheorem{rem-eg}[thm]{Remark and Example}

\SetKw{KwInitialization}{Initialization:}

\begin{document}
\title{Distributed Storage Schemes over Unidirectional Ring Networks}

\author{
  \IEEEauthorblockN{Jiyong~Lu}
  \IEEEauthorblockA{Chern Institute of Mathematics\\
    Nankai University\\
    Tianjin, P. R. China\\
    Email: lujiyong@mail.nankai.edu.cn}
  \and
  \IEEEauthorblockN{Xuan~Guang}
  \IEEEauthorblockA{School of Mathematical Science\\
    Nankai University\\
    Tianjin, P. R. China\\
    Email: xguang@nankai.edu.cn}
  \and
  \IEEEauthorblockN{Fang-Wei~Fu}
  \IEEEauthorblockA{Chern Institute of Mathematics and LPMC\\
    Nankai University\\
    Tianjin, P. R. China\\
    Email: fwfu@nankai.edu.cn}}

\markboth{Distributed Storage over Unidirectional Ring Networks}%
{\MakeLowercase{\textit{et al.}}: Distributed Storage over Unidirectional Ring Networks}

\maketitle

\begin{abstract}
In this paper, we study distributed storage problems over unidirectional ring networks. A lower bound on the reconstructing bandwidth to recover total original data for each user is proposed, and it is achievable for arbitrary parameters. If a distributed storage scheme can achieve this lower bound with equality for each user, we say it an optimal reconstructing distributed storage scheme (ORDSS). Furthermore, the repair problem for a failed storage node in ORDSSes is under consideration and a tight lower bound on the repair bandwidth for each storage node is obtained. Particularly, we indicate the fact that for any ORDSS, every storage node can be repaired with repair bandwidth achieving the lower bound with equality. In addition, we present an efficient approach to construct ORDSSes for arbitrary parameters by using the concept of Euclidean division. Finally, we take an example to characterize the above approach.
\end{abstract}

%
\IEEEpeerreviewmaketitle

\section{Introduction}

\IEEEPARstart{D}{istributed} storage schemes for data storage have been studied for a long period since they can keep data reliable over unreliable nodes. All kinds of strategies are proposed for data storage, such as replication \cite{Bolosky-etc-2000}, erasure cods \cite{Xu-etc-1999}, and regenerating codes \cite{Dimakis-etc-2010}, etc. Among them, regenerating codes recently proposed by Dimakis \textit{et al.} \cite{Dimakis-etc-2010} are more effective in terms of repair bandwidth. Motivating by network coding \cite{L-Y-C}, they used an information flow graph to express regenerating codes, and identified a tradeoff curve between the storage capacity per node and the repair bandwidth for a failed node. This tradeoff curve has two extremal points which correspond to the minimum storage regenerating (MSR) codes \cite{Cullina-Dimakis-Ho-2009}, \cite{Rashmi-Shah-Kumar-2011}, and the minimum bandwidth regenerating (MBR) codes \cite{Rashmi-Shah-Kumar-2011}, respectively. However, lots of distributed storage schemes do not consider network structures among storage nodes. Actually, in many applications, storage nodes have certain topological relationships, such as the hierarchical network structure \cite{Benson-Akella-Maltz-2010}, the multi-hop network structure \cite{Kong-Aly-Soljanin-2010} and so on. Li \textit{et al.} \cite{Li-etc-2010} studied repair-time in tree-structure networks which have links with different capacities. In \cite{Shum-etc}, Gerami \textit{et al.} studied repair-cost in multi-hop networks and formulated the minimum-cost as a linear programming problem for linear costs.

Inspired by these works, in this paper, we focus on distributed storage problems over a class of simple but important networks, i. e., unidirectional ring networks, which usually exist as parts of some complex networks. In these unidirectional ring networks, each user connects one and only one storage node to download data. For each user, its reconstructing bandwidth is the total number of all transmitted symbols to recover original data. By cut-set bound analysis of information flow graph for each user requiring total original data in this system, we obtain a lower bound on the reconstructing bandwidth and further indicate its tightness for arbitrary parameters. Thus, we define optimal reconstructing distributed storage schemes (ORDSSes), if the reconstructing bandwidth for every user achieves the lower bound with equality. Furthermore, we study the repair problem in ORDSSes and also deduce a tight lower bound on the repair bandwidth for repairing each failed storage node, which is the total number of all transmitted symbols to repair this failed storage node. In particular, we show that every ORDSS can satisfy the lower bound with equality. At last, we present two constructions of ORDSSes, called MDS construction and ED construction, respectively. Both of them can be applied for arbitrary parameters, but ED construction has many advantages than MDS construction, such as finite field size and computational complexity.

The remainder of the paper is organized as follows. In Section \ref{M-Example}, a motivating example is given to illustrate our research problems. The basic mathematical model of research problems and tight lower bounds on the reconstructing and repair bandwidths are deduced in Section \ref{BM-SBound}. In Section \ref{cons}, we design an efficient constructing approach of ORDSSes for arbitrary parameters $(n,\alpha,M)$. Finally, Section \ref{conc} concludes this paper.
Because the limit of the space, some proofs of our conclusions are ignored in this paper. For more details, please refer to the extended reversion \cite{L-G-F}.

\section{A Motivating Example }\label{M-Example}
\begin{figure}[!ht]
\centering
\includegraphics[width=8.1cm,height=4.2cm]{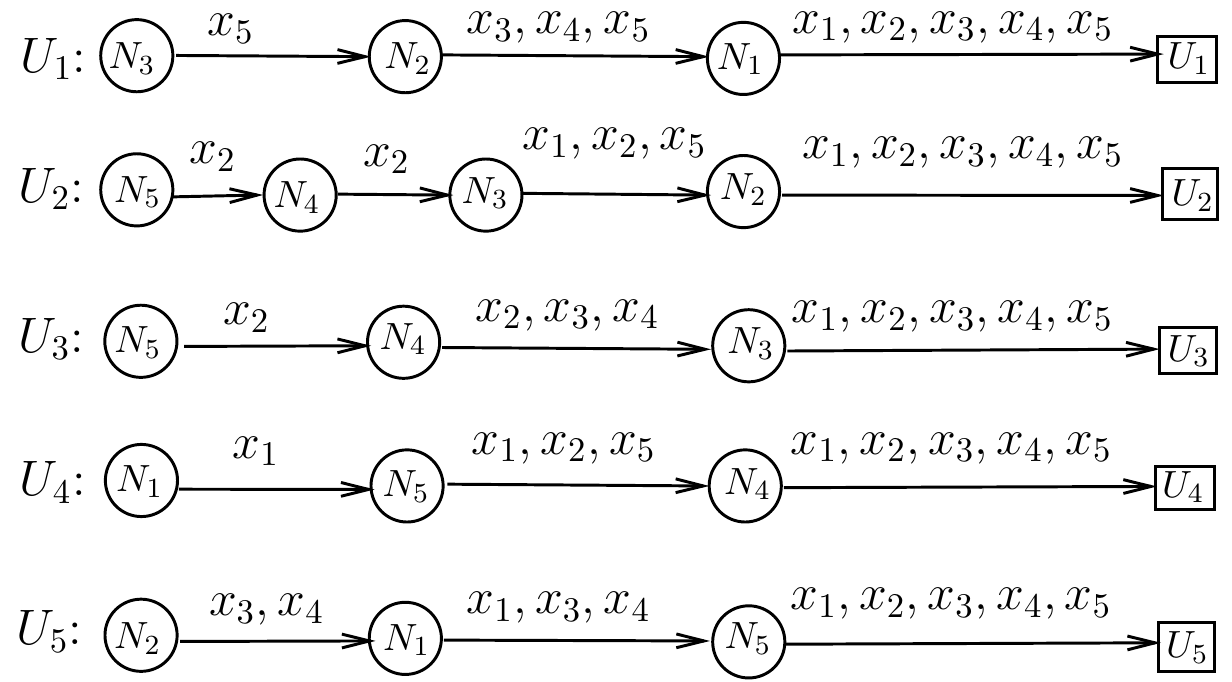}
\caption{The reconstructing process for all users. Each of $U_1,U_3,U_4$ obtains the original data with bandwidth 9, while each of $U_2,U_5$ with bandwidth 10.}
\label{fig_rc}
\end{figure}

In this section, we first discuss an example to show our research problems. This shows that it is meaningful and interesting to study the distributed storage problems over unidirectional ring networks.
\begin{figure}[!ht]
\centering
\includegraphics[width=4.8cm,height=3.1cm]{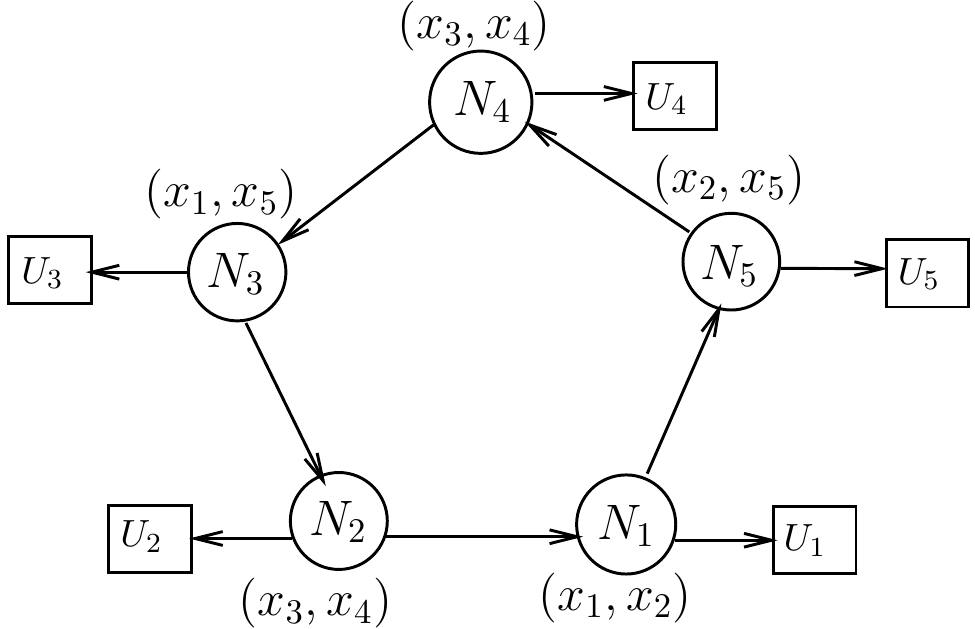}
\caption{The unidirectional ring network with $n=5$, $\alpha=2$, $M=5$.}
\label{fig_cn}
\end{figure}
Fig. \ref{fig_cn} depicts a unidirectional ring network with $n=5$ storage nodes, denoted by $N_1,N_2,\cdots,N_5$, the data exchanges between the storage nodes have to accord to the direction of the ring network. Each storage node has storage capacity $\alpha=2$. Let the row vector of original data be $X=[x_1,x_2,x_3,x_4,x_5]\in\mathbb{F}^5_5$, that is, the size $M$ of original data is 5. All five storage nodes distributed store total original data $X$. Each user connects one and only one storage node in order to download total original data. Without loss of generality, let user node $U_i$ connect storage node $N_i$, $1\leq i\leq 5$. Fig. \ref{fig_cn} also gives a distributed storage scheme, in which every user can download the original data. For instance, $N_1$ stores $x_1$ and $x_2$, $N_2$ stores $x_3$ and $x_4$. For this scheme, Fig. \ref{fig_rc} describes an optimal reconstructing process which minimizes the reconstructing bandwidth for every user.
Based on this reconstructing process, we know that the minimum reconstructing bandwidth on average of this scheme is $47/5=9.4$. Naturally, we propose a series of problems as follows: does there exist distributed storage schemes with the minor reconstructing bandwidth? what is the minimum of the reconstructing bandwidth? how to efficiently construct distributed storage schemes achieving the minimum reconstructing bandwidth?

\begin{figure}[!h]
\begin{center}
\includegraphics[width=4.8cm,height=3.2cm]{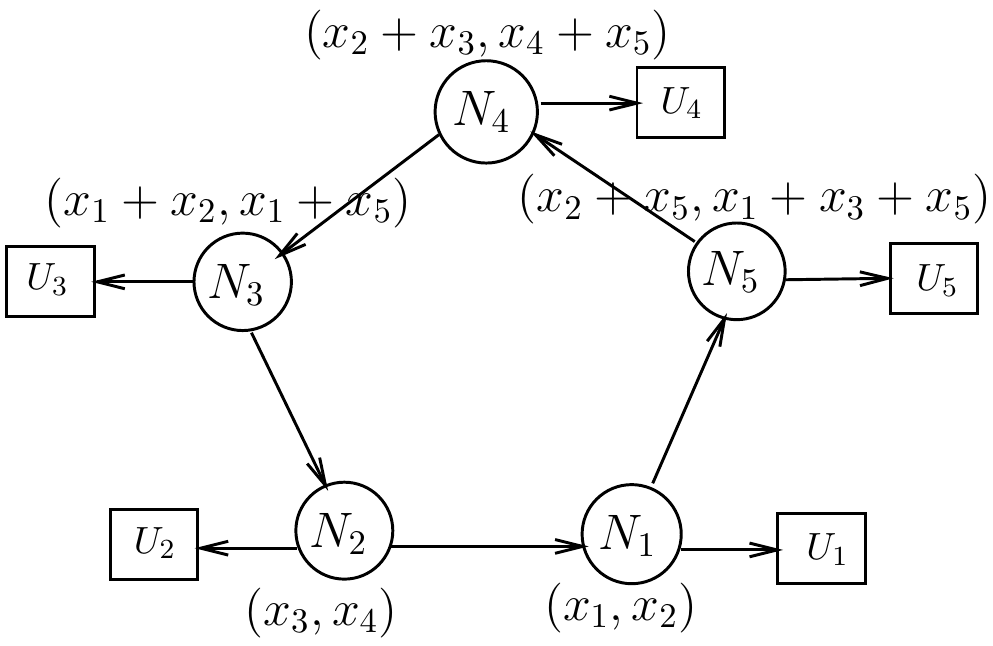}
\caption{The new storage scheme with $n=5,\alpha=2,M=5$.}
\label{fig_mrc}
\end{center}
\end{figure}
\begin{figure}[!h]
\begin{center}
\includegraphics[width=8cm,height=4cm]{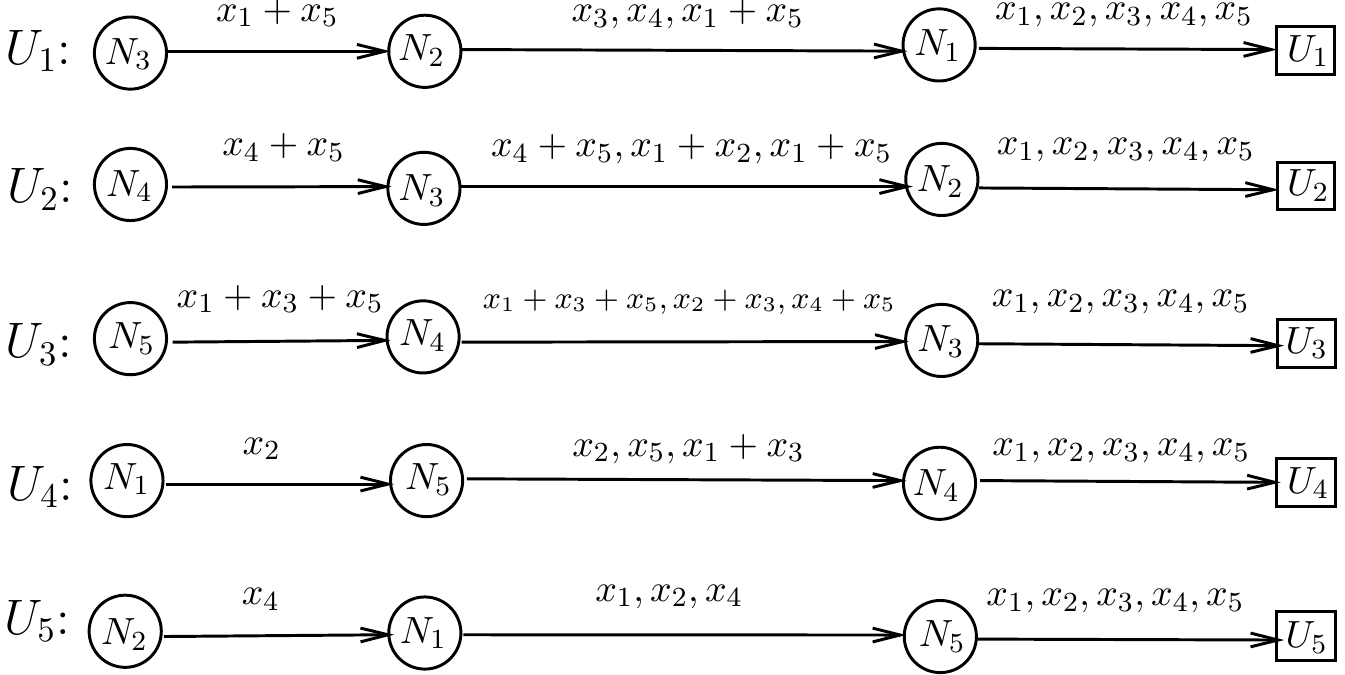}
\caption{The reconstructing process of the new storage scheme. Each user downloads the original data with reconstructing bandwidth 9.}
\label{fig_mrp}
\end{center}
\end{figure}

In fact, the above distributed storage scheme is not optimal, there exists a better storage scheme as described in Fig. \ref{fig_mrc}, where each user consumes reconstructing bandwidth 9 to download the original data.
Fig. \ref{fig_mrp} characterizes its reconstructing process. Thus, the average minimum reconstructing bandwidth of this new scheme is 9. Actually, in this example, 9 is the minimum reconstructing bandwidth for any user.

\begin{figure}[!h]
\begin{center}
\includegraphics[width=8cm,height=4cm]{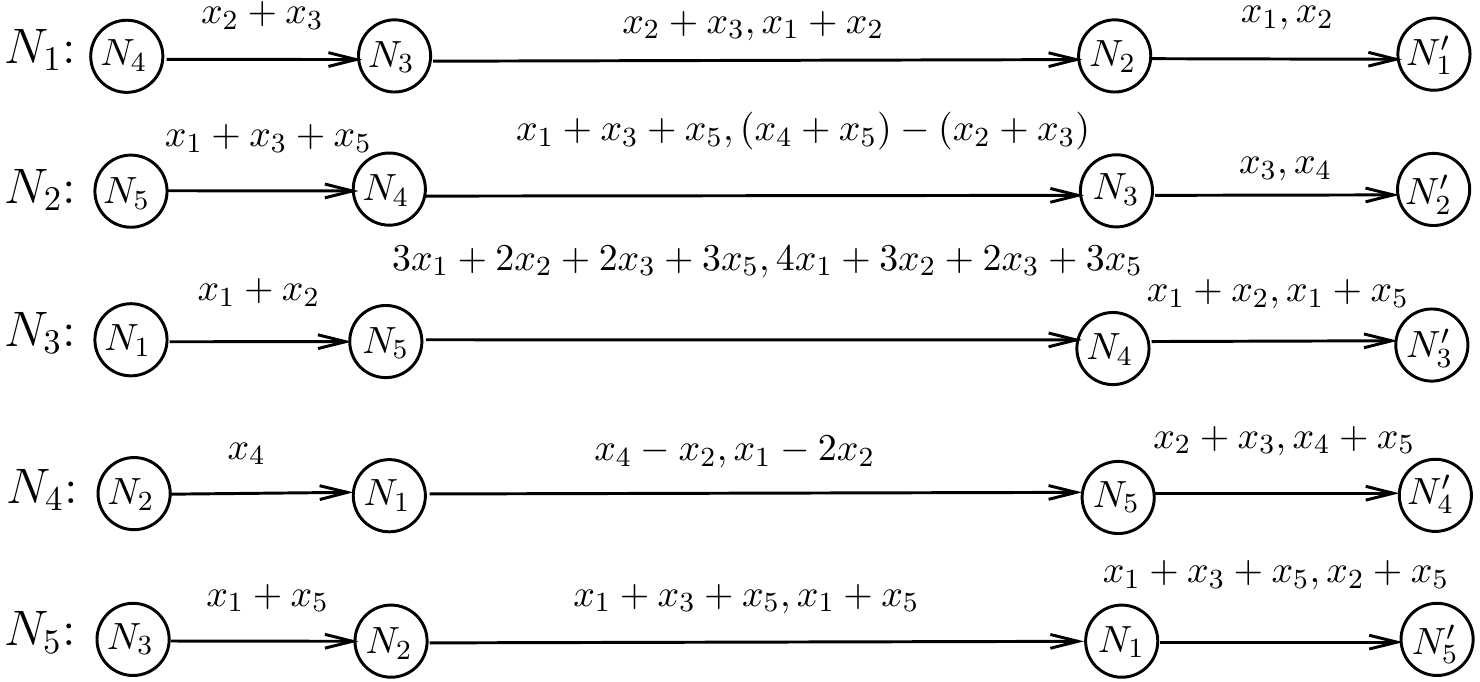}
\caption{The optimal repair process of the new storage scheme, and the average minimum repair bandwidth of this scheme is 5.}
\label{fig_mrpair}
\end{center}
\end{figure}

Further, if some storage node fails, several interesting and meaningful problems should be considered. For instances, can this failed node be repaired? what is the minimum repair bandwidth? how to repair it efficiently? Fig. \ref{fig_mrpair} characterizes the optimal repair process with the minimum bandwidth 5 for each storage node in the new storage scheme as depicted in Fig. \ref{fig_mrc}.

\section{Basic Model and Some Bounds}\label{BM-SBound}

In this section, we will describe the basic model of our research problems over unidirectional ring networks. The lower bounds on the reconstructing and repair bandwidths are also proposed.

\subsection{Basic Model}
Let $\mathcal {G}$ be a unidirectional ring network consisting of $n$ storage nodes, denoted by $N_1,N_2,\cdots, N_n$, and each one has a capacity to store $\alpha$ symbols. These $n$ storage nodes form a directed ring and data is only transmitted according to the given direction. Let $X=[x_1,x_2,\cdots,x_M]$ be the row vector of original data with size $M$, each coordinate represents an information symbol taking values in a finite field $\mathbb{F}_q$ with $q$ elements, where $q$ is a power of some prime. The data $X$ is distributed to all storage nodes in order to store $X$. Here, we just consider linear storage, that is, every stored symbol and every transmitted symbol are linear combinations of the information symbols, still being an element in $\mathbb{F}_q$.

For any storage node $N_i$, define an $M \times \alpha$ node generator matrix $G^{(i)}$. Then all the $\alpha$ coordinates of the product $X G^{(i)}$ are stored in $N_i$, each of which is called a node symbol. And each node symbol corresponds to a column vector of $G^{(i)}$, called node vector. Further, each transmitted symbol is a linear combination of some node symbols, and clearly, it also corresponds to a vector, called a transmitted vector, which is the linear combination of those corresponding node vectors, too. Concatenating all node generator matrices according to the order of storage nodes, we obtain an $M \times n \alpha$ matrix  $G=[G^{(1)},G^{(2)},\cdots,G^{(n)}]$, which is called a generator matrix of this distributed storage scheme. Each user connects one and only one storage node to download data. Note that, in order to ensure all users to reconstruct the original data completely in this ring network, the generator matrix has to be full row-rank.

We apply information flow graph to analyze the following reconstructing bandwidth, which is a particular graphical representation of distributed storage schemes.

\begin{figure}[!ht]
\centering
\includegraphics[width=6.5cm,height=4cm]{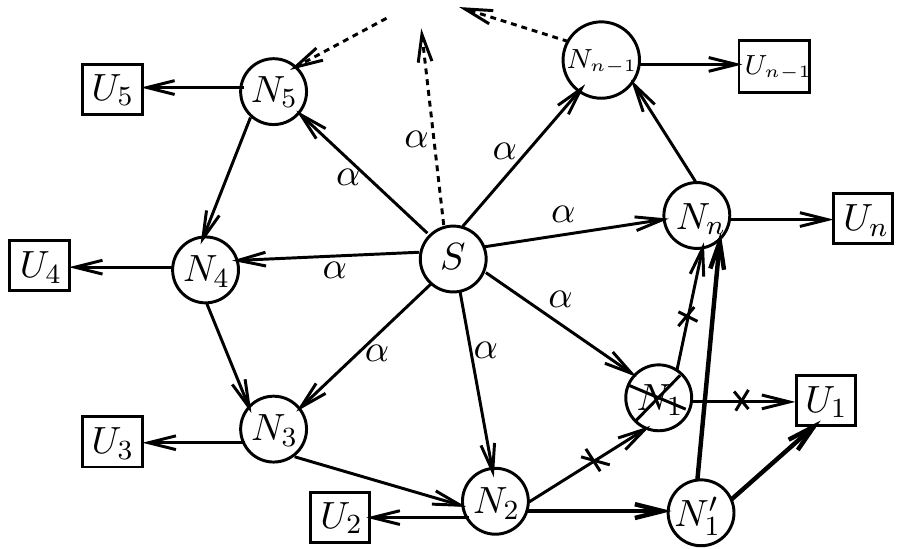}
\caption{The information flow graph $\mathcal {G}$, where circular nodes and rectangular nodes represent the storage nodes and the user nodes, respectively. When the storage node $N_1$ fails, a new node $N^\prime_{1}$ arises and establishes connections from the node $N_2$, to the node $N_n$ and the user node $U_1$.}
\label{inf-graph}
\end{figure}

$Information-Flow-Graph$: An information flow graph $\mathcal {G}$ consists of three types of nodes: a single source node $S$ (the source of original data), $n$ storage nodes and some user nodes. The source node $S$ connects to $n$ storage nodes with directed edges with capacity $\alpha$. After the source node $S$  distributes node symbols, it becomes inactive. All storage nodes form a directed ring through edges with capacity $M$. The edges connecting user nodes and their corresponding storage nodes also have capacity $M$. When some storage node $N_i$ fails, $1\leq i\leq n$, a new substituted node $N^\prime_{i}$ arises to replace it and establishes connections from the node $N_{i+1}$, to the node $N_{i-1}$ and the users. Due to the symmetry, we just take one user connecting to the same  storage node into account. Fig. \ref{inf-graph} indicates the details of the information flow graph $\mathcal {G}$.

In the following, we introduce some concepts in graph theory which are used in the paper.
A cut in the graph $\mathcal {G}$ between the source node $S$ and a fixed user $U$ is a subset of edges whose removal  disconnects $S$ from $U$. The minimum cut  between $S$ and $U$ is a cut between them in which the total sum of the edge capacities achieves the smallest.

\subsection{Bounds on Reconstructing and Repair Bandwidths}
For unidirectional ring networks, there are three important parameters: $n$, the number of storage nodes; $\alpha$, the storage capacity per storage node; and $M$, the size of original data.
\begin{thm}\label{thm-rec-bound}
 For any storage scheme of a unidirectional ring network with parameters $(n,\alpha,M)$, the reconstructing bandwidth to recover the original data for each user is lower bounded by $kM-\frac{(k-1)k\alpha}{2}$, where $k=\lceil M/\alpha\rceil$. Moreover, there exists a storage scheme such that all users can reconstruct original data with reconstructing bandwidth achieving this lower bound with equality.
\end{thm}

Before the proof of Theorem \ref{thm-rec-bound}, we need the following two lemmas firstly.
\begin{lemma}\label{lemma-min-cut}{\cite[Lemma 1]{Dimakis-etc-2010}}
No user $U$ can reconstruct the original data if the minimum cut capacity in a directed acyclic graph between the source node $S$ and $U$ is smaller than the original data size $M$.
\end{lemma}

\begin{lemma}\label{lemma-iff}
For a storage scheme over a unidirectional ring network with parameters $(n,\alpha,M)$, all users can recover the original data with the same reconstructing bandwidth $kM-\frac{(k-1)k\alpha}{2}$, where $k=\lceil M/\alpha\rceil$, if and only if the following two conditions are satisfied:

(i) all $(k-1)\alpha$ node vectors of arbitrary $k-1$ adjacent storage nodes are linearly independent;

(ii) arbitrary $k$ adjacent storage nodes contain $M$ linearly independent node vectors.
\end{lemma}

\begin{IEEEproof}[Proof of Theorem \ref{thm-rec-bound}]
Because of the symmetry of this network, it suffices to consider the reconstructing bandwidth for any one user. Without loss of generality, we take the user $U_1$ into account, which is connected from the storage node $N_1$ as depicted in Fig. \ref{inf-graph}. Clearly, when $U_1$ is under consideration, it is not necessary to transmit data from $N_1$ to $N_n$. So we can omit the edge from $N_1$ to $N_n$ and only consider the degenerated acyclic graph $\mathcal {H}$ of the cyclic graph $\mathcal {G}$ as depicted in the following Fig. \ref{fig_ag}.
\begin{figure}[!ht]
\centering
\includegraphics[width=6.5cm,height=4.2cm]{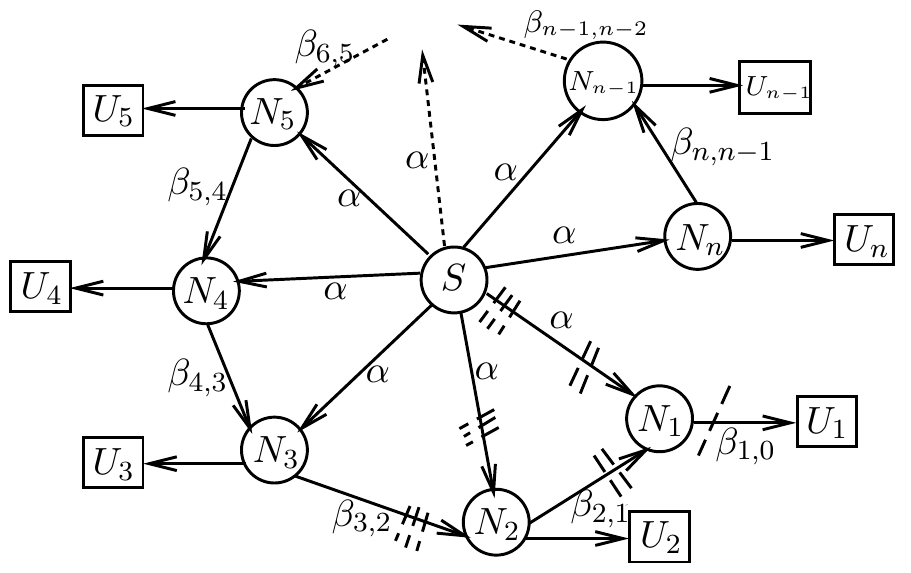}
\caption{The degenerated acyclic graph $\mathcal {H}$.}
\label{fig_ag}
\end{figure}
In the acyclic graph $\mathcal {H}$, let $\beta_{i,i-1}$ be the amount of transmitted vectors from $N_i$ to $N_{i-1}$, where $1\leq i \leq n$, and $N_0$ represents the user node $U_1$ in order to keep consistency of notation. Thus, the total reconstructing bandwidth for $U_1$ is $\sum^{n}_{i=1}\beta_{i,i-1}$. Next, we will propose an lower bound by cut-set bound analysis method.

By Lemma \ref{lemma-min-cut}, the minimum cut capacity between $S$ and $U_1$ should not be less than $M$, which implies that we just need to analyze those  potential minimum cuts between $S$ and $U_1$. In the following, we list all possibilities:

\begin{equation*}
\begin{aligned}
\beta_{1,0} & \geq M, \\
\alpha+\beta_{2,1} & \geq M, \\
\cdots\\
(n-1)\alpha+\beta_{n,n-1} & \geq M.
\end{aligned}
\end{equation*}
Actually, we have $k\alpha\geq M$ as $k=\lceil M/\alpha\rceil$, which implies the last $(n-k)$ inequalities are useless. Hence, the above inequalities imply the lower bound below:
\begin{align*}
\sum\limits^{n}_{i=1}\beta_{i,i-1} & \geq M+(M-\alpha)+\cdots+[M-(k-1)\alpha]\\
&=kM-\frac{(k-1)k\alpha}{2}.
\end{align*}

Next, we consider the tightness of this bound. Let $G$ be a generator matrix of an $[n\alpha,M]$ MDS code, and we regard it as the generator matrix for our storage scheme. Then we partition all $n\alpha$ column vectors into $n$ parts, each of which contains $\alpha$ column vectors constituting node generator matrix for a storage node. It is easy to check that this storage scheme satisfies the two conditions in Lemma \ref{lemma-iff}. Thus, all users can reconstruct the original data with bandwidth $kM-\frac{(k-1)k\alpha}{2}$, which implies that the proposed lower bound is achievable for all users. This completes the proof.
\end{IEEEproof}

If a distributed storage scheme achieves the lower bound in Theorem \ref{thm-rec-bound} with equality for all users, we say it an optimal reconstructing distributed storage scheme (ORDSS). Actually, any $[n\alpha, M]$ MDS code can constitute an ORDSS. Particularly, by the definition of ORDSS, the two conditions in Lemma \ref{lemma-iff} are actually sufficient and necessary for an ORDSS.

In an ORDSS, when a storage node fails, a new node arises to replace it. Next, we will focus on the repair bandwidth in order to repair this failed storage node in any ORDSS.
\begin{thm}\label{thm-repair}
For any ORDSS over a unidirectional ring network with parameters $(n,\alpha,M)$, the repair bandwidth for any failed storage node is lower bounded by $M$.
\end{thm}

\begin{IEEEproof}
The sketch of the proof is presented below, for more details, please refer to the extended version \cite{L-G-F}. Different from the proof in Theorem \ref{thm-rec-bound}, we mainly apply linear algebra as analysis technique in this proof.

We only take the repair of the storage node $N_1$ into account as the symmetry of the network. Let $M=(k-1)\alpha+\gamma$, where $k=\lceil M/\alpha \rceil$ and $0<\gamma\leq\alpha$.

Let $\eta_{i,i-1}$ be the number of transmitted vectors from $N_i$ to $N_{i-1}$ for repairing $N_1$, $2\leq i\leq n$. By Lemma \ref{lemma-iff} and the linear relationship of node vectors in storage nodes, we can deduce that $\eta_{i,i-1}\geq \alpha$ for $2\leq i\leq k$ and $\eta_{k+1,k}\geq\gamma$. Therefore, the storage nodes $N_2,\cdots,N_{k+1}$ have to provide at least $\sum^{k+1}_{i=2}\eta_{i,i-1}=(k-1)\alpha+\gamma=M$ vectors to repair $N_1$. Thus, $M$ is a lower bound on the repair bandwidth.
\end{IEEEproof}

For the above lower bound on the repair bandwidth, wether there exist an ORDSS to achieve it for all storage nodes. The following theorem answers this question.
\begin{thm}\label{thm-ORDSS-repair}
For any ORDSS over a unidirectional ring network with parameters $(n,\alpha,M)$, any storage node can be repaired successfully with repair bandwidth $M$.
\end{thm}

\section{Construction of ORDSSes}\label{cons}
According to Theorem \ref{thm-ORDSS-repair}, any ORDSS can achieve the lower bound on the repair bandwidth with equality in Theorem \ref{thm-repair}. Thus, it is sufficient to construct ORDSSes, and in the following, we use the concept of Euclidean division to present an efficient construction of ORDSSes.

For any two finite positive integers $N$ and $M_0$ with $N\geq M_0$, by Euclidean division, we have the following equalities for some integer $k$:
\begin{equation*}\left\{
\begin{array}{ll}
N=P_0M_0+M_1, & 0\leq M_1<M_0\\
M_0=P_1M_1+M_2, & 0\leq M_2<M_1\\
\cdots & \cdots\\
M_{k-1}=P_{k}M_{k}+M_{k+1},& 0\leq M_{k+1}<M_k\\
M_k=P_{k+1}M_{k+1}.&
\end{array}\right.
\end{equation*}
Note that $M_{k+1}={\rm gcd}(N,M_0)$.

\begin{defn}\label{def3}
Define an $M_0\times N$ matrix $G$ as follows:

\begin{equation*}
\left[
\begin{array}{ccc@{}|c@{}}
I_{M_0} & \cdots & I_{M_0} & \begin{array}{c}
I_{M_1}\\
\vdots\\
I_{M_1}\\\hline
\begin{array}{ccc|c}
I_{M_2} & \cdots & I_{M_2} & \begin{array}{c}
I_{M_3}\\
\vdots\\
I_{M_3}\\\hline
\cdots
\end{array}
\end{array}
\end{array}
\end{array}\right],
\end{equation*}
where $I_{M_i}$ represents $M_i\times M_i$ identity matrix and the number of $I_{M_i}$ is $P_i$, $0 \leq i \leq k+1$. This matrix $G$ is called ED-matrix.
\end{defn}

\begin{defn}\label{def-weakly MDS property}
For an $M \times N$ matrix, it is said to satisfy weakly MDS property, if its arbitrary cyclic adjacent $M$ columns (or, $N$ rows) are linearly independent when $M\leq N$ (or, $M>N$). Here, ``cyclic adjacent'' means that the last column (resp. row) of this matrix is regarded to be adjacent with the first column (resp. row).
\end{defn}

\begin{thm}\label{thm-MDS}
ED-matrices satisfy the weakly MDS property.
\end{thm}

In the following, we present an efficient construction of ORDSSes.
For a unidirectional ring network with arbitrary parameters $(n,\alpha,M)$, we select an $M\times n\alpha$ ED-matrix $G$ as the generator matrix of the distributed storage scheme. Similarly, $X$ is the $M$-dimensional row vector of original data. Then assign the $n\alpha$ coordinates of the product $XG$ (equivalently, the $n\alpha$ column vectors of $G$) to $n$ storage nodes as following approach: assign the first $\alpha$ coordinates to the storage node $N_1$, the second $\alpha$ coordinates to the storage node $N_2$, so far and so forth, the last $\alpha$ coordinates, i.e., the $n$th $\alpha$ coordinates, to the storage node $N_n$.

By Theorem \ref{thm-MDS}, we know that arbitrary $M$ cyclic adjacent columns of ED-matrix $G$ are linearly independent, that is, one can obtain the original data from arbitrary $M$ cyclic adjacent coordinates of $XG$. Thus, the above assignment shows that the node vectors in arbitrary $k-1$ adjacent storage nodes are linearly independent and arbitrary $k$ adjacent storage nodes contain $M$ linearly independent vectors. Together with Lemma \ref{lemma-iff}, this storage scheme is an ORDSS. Therefore, all users can recover the original data with the minimum reconstructing bandwidth $kM-\frac{(k-1)k\alpha}{2}$. In addition, Theorem \ref{thm-ORDSS-repair} shows that any storage node can be repaired with bandwidth $M$ if it fails.

\begin{eg}\label{EX_3}
For a unidirectional ring network with parameters $(n=4,\alpha=2,M=5)$, let $X=[x_1, x_2, x_3, x_4, x_5]\in \mathbb{F}^5_2$ be the row vector of original data. We construct an ED-matrix $G$ of size $5 \times 8$ as follows:

$$G=\left[\begin{array}{cccccccc}1&0&0&0&0&1&0&0\\0&1&0&0&0&0&1&0\\
0&0&1&0&0&0&0&1\\0&0&0&1&0&1&0&1\\0&0&0&0&1&0&1&1\end{array}\right].$$
Subsequently, we calculate $$XG =[x_1, x_2, x_3, x_4, x_5, x_1+x_4, x_2+x_5, x_3+x_4+x_5].$$
Then, we assign $x_1,x_2$ to the node $N_1$, $x_3,x_4$ to the node $N_2$, $x_5, x_1+x_4$ to the node $N_3$ and $x_2+x_5$, $x_3+x_4+x_5$ to the node $N_4$.
Clearly, any user can reconstruct the original data $X$ with reconstructing bandwidth 9. And if any storage node fails, it can be repaired with repair bandwidth 5. For instance, if the node $N_2$ fails, $N_1$ transmits $x_1$ to $N_4$, then $N_4$ transmits $x_1,x_3+x_4+x_5$ to $N_3$, $N_3$ can solve $x_3,x_4$ and transmits them to the new substituted node $N^\prime_{2}$. So $N_2$ is repaired exactly with repair bandwidth 5. Fig. \ref{ED-construction} depicts the repair process in details for this storage scheme.
\begin{figure}[!ht]
\centering
\includegraphics[width=8.4cm,height=3.2cm]{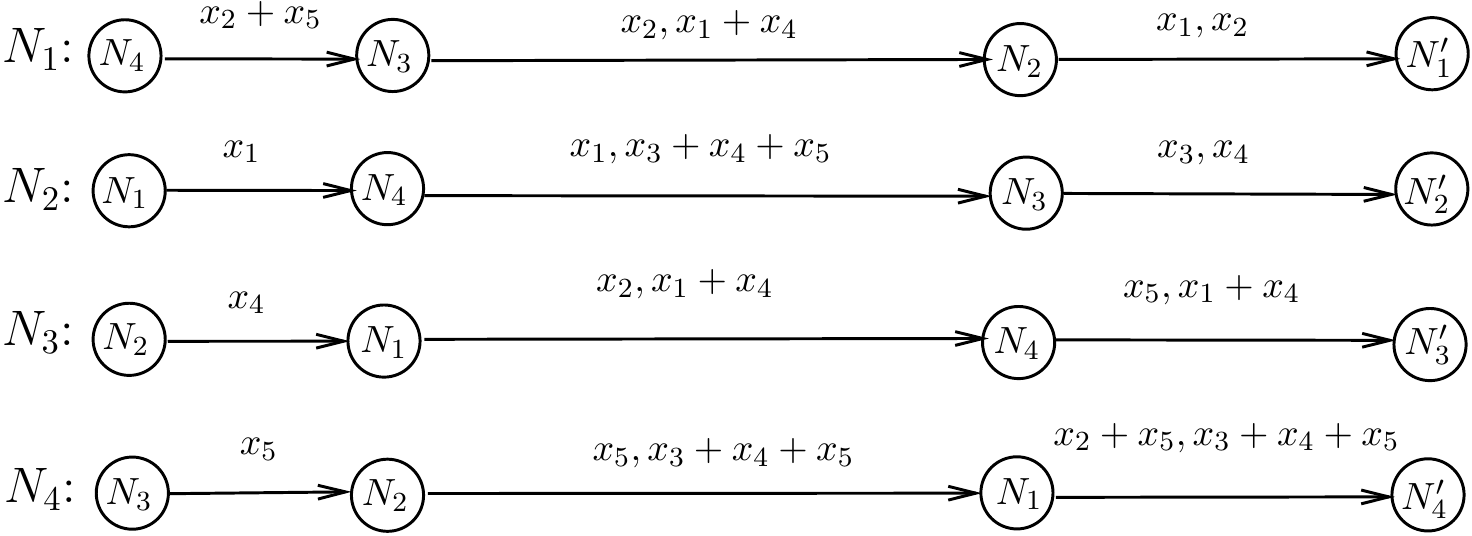}
\caption{The repair process for this storage scheme. }
\label{ED-construction}
\end{figure}
\end{eg}

We call the construction of ORDSSes in the proof of Theorem \ref{thm-rec-bound} the MDS construction, and the above construction using an ED-matrix the ED construction. Now, we compare the two constructions. MDS construction applies a generator matrix $G$ of an $[n\alpha, M]$ MDS code, whose arbitrary $M$ columns are linearly independent. This property, called MDS property, is too strong for constructing ORDSSes. And the authors in \cite{Xiao-Qing-1993} have shown that the size of finite field $\mathbb{F}_q$ is no less than $n-k+1$ for the existence of an $[n,k]$ MDS code with $k\geq 2$. For MDS construction, when the values of parameters $(n,\alpha,M)$ are large, the field size will become too large to be applied in practice, while ED construction always uses the smallest finite field $\mathbb{F}_2$. This is because that the weekly MDS property is sufficient for constructing ORDSSes. For example, when $n=500,\alpha=10,M=1000$, the field size for a $[5000,1000]$ MDS construction is at least $5000-1000+1=4001$, which is much larger than the field size 2 for ED construction. It is well-known that the cost of arithmetic in a small field is smaller than that in a bigger one. Thus, the smaller field size will reduce the computational complexity of the storage scheme and save much time evidently. Therefore, ED construction is much better than MDS construction.

\section{Conclusion}\label{conc}
In this paper, we discuss distributed storage problems over unidirectional ring networks and propose two tight lower bounds on the reconstructing and repair bandwidths. In addition, we present two constructions for ORDSSes, and both can be used for arbitrary parameters. In practical applications, the networks of bidirectional ring topology, in which adjacent nodes can exchange information each other, are more useful. The same research problems in that case are also meaningful and still keep open.

\section{Acknowledge}
This research is supported by National Key Basic Research Problem of China (973 Program Grant No. 2013CB834204), the National Natural Science Foundation of China (Nos. 61301137, 61171082) and Fundamental Research Funds for Central Universities of China (No. 65121007).


\end{document}